# Fermi surface topology and magneto-quantum oscillations for under-doped YBCO with hopping anisotropy

**Partha Goswami**

Physics department, D.B.College, University of Delhi, Kalkaji, New Delhi-110019, India

**PACS** 74.20.-z ,74.72.-h, 74.20.Rp, 74.72.Bk,74.20.De.

Corresponding author: e-mail physicsgoswami@gmail.com, Phone: +91 129 243 90 99.

We investigate a chiral d-density wave (CDDW) mean field Hamiltonian in momentum space, which also includes the in-plane hopping anisotropy parameter έ, for the under-doped $YBa_2Cu_3O_{7-\delta}$ (YBCO) to explore the possibility of quantum oscillations (QO) in the specific heat in the presence of a changing magnetic field (B). The inclusion is motivated by the experimental signature of nematic order found in cuprates in neutron scattering experiments of Hinkov et al. (see Ref.[10]). The anisotropy is found to have negligible effect on the location of the peaks in the momentum dependent coefficient of the transverse heat current for B = 0 which carries the signature of chirality. We obtain the Fermi surface topologies, without and with magnetic field background, including the elastic scattering by impurities in the investigation within Γ-matrix approximation framework. The topologies are found to be distinct from the Fermi arc picture: we find that the minimally gapped portion of Fermi surface forms closed loops (hole pockets) and not arcs. However, some **k**-points of the electron pockets are found to be associated with the negative density of state values (see Ref.[28]) when B = 0 and έ = 0; for B ≠ 0 and έ → $0^+$ these pockets acquire positive DOS values. We relate our findings regarding the Fermi surface topology to QO in specific heat. We show that the origin of the main frequency of the oscillations is the electron pocket of the Fermi surface at a magnetic field B ~ 50 T.

**1 Introduction** The pseudo-gapped phase of YBCO has been characterized with a variety of coexisting or competing orders by previous workers[1,2,3,4,5,6,7,8,9,10,11,12]. This has led to different Fermi surface topologies and anomalies [13,14,15,16,17,18,19] in many physical properties. In this communication, starting with a chiral d-density wave (CDDW)[1,2] mean field Hamiltonian $H_{d+id}$ in momentum space involving in-plane hopping anisotropy for PG state of under-doped YBCO[10], we show that the approximate (1/B)-oscillations in specific heat, similar to the anomalies in the conductivity and the magnetization [14,15,16,17,18,19], is possible. The main frequency of the specific heat oscillations is found to have its origin in the electron pocket of the Fermi surface(FS) obtained from the energy eigenvalues of the matrix in $H_{d+id}$. The reason for the identification of the PG state with the CDDW state rather than the well-known [3,4] d-density wave (DDW) state is that the CDDW ordering offers a theoretical explanation [5] of the non-zero polar Kerr effect observed recently in YBCO by Kapitulnik et al.[11].The inclusion of the elastic scattering by impurities in Γ-matrix approximation, without and with magnetic field background, leads to Fermi surface topologies distinct from the Fermi arc picture: we find that the minimally gapped portion of Fermi surface forms closed loops (hole pockets) and not arcs. The reason is that the coherence factors appearing in the expression of the ensemble averaged full Matsubara propagator, arising out of the Born approximation followed by the Γ-matix approximation in our investigation, are essentially complex functions. The expression of the Fermi energy density of states(DOS) $\rho_{Fermi}(k)$ (or spectral function) therefore involves, apart from the bunch of Lorenztians multiplied with the real part of the coherence factors, additional terms including imaginary parts. In the absence of the latter, one obtains the usual Fermi arc feature of the previous theoretical studies[3,4,7,8] in $\rho_{Fermi}(k)$ while the presence leads to the closed loops alluded to above. It must be added that, unlike the aforementioned studies, the "algebraic charge liquid" picture of Senthil et al. [9] predicts two kinds of hole-like Fermi pockets, viz. the elliptic and the banana-shaped. Our finding is consistent with the former and not with the latter. The angle resolved photo-emission spectroscpic(ARPES) studies [20,21,22,23,24,25](including the vacuum ultraviolet (VUV) laser-based ARPES [26]), where the experimental observations roughly correspond to the so-called "maximal intensity surface" explained in ref.[22], however have not shown the evidence of the existence of the Fermi pockets so far.

The real motivation behind our investigation of the specific heat anomaly (SHA) is the experimental finding of G. S. Boebinger[13] and his collaborators. They have observed quantum oscillations in the specific heat of YBCO-Ortho II samples, the same type YBCO samples investigated by Proust, Taillefer and co-workers [14,15,16] who recently detected quantum oscillations in the electrical resistance of under-doped YBCO establishing the existence of a well-defined Fermi surface with Fermi pockets when the superconductivity is suppressed by a magnetic field. We wish to mention that in a previous paper [27], hereinafter referred to as I, we have gone ahead with the problem of SHA pending the inclusion of the scattering by imperfections. In this communication we have taken up that unfinished task. Since the experimental signature of nematic order has been observed recently in cuprates in neutron scattering experiments (see Ref.[10]), it is felt that the hopping anisotropy (the corresponding parameter is denoted by $\acute{\varepsilon}$ ) ought to be an integral part of the PG state investigation of SHA (involving disorder). The nature of the transition from the normal metallic phase to the CDDW state has been ascertained in I: We reported that at certain spots ('Hot spots') on the boundary of the reduced Brillouin zone(RBZ), $k_x \pm k_y = \pm\pi/a$, close to the Fermi pockets there are jump discontinuities in the entropy density difference (EDD) $\Delta s(\mathbf{k})$ between the CDDW state and the normal metallic state; on the boundaries of the electron and hole pockets EDD have been found to peak dramatically. These spots are found to be intimately linked with the transition to the CDDW ordered phase, as without these spots the $\sum\Delta s(\mathbf{k})$ would become positive. It was thus concluded that the transition is first order. For the chirality induced anomalous Nernst signal (ANS) which involves the entropy in the CDDW state, we found in I that the peaks were located at the points common to the RBZ boundary and the hole pockets in the momentum space as reported by the previous workers [1].We find here, quite surprisingly, that the inclusion of the hopping anisotropy [10] has negligible effect on the ANS peaks indicating that the chirality is in no competition with the anisotropy in the under-doped YBCO.

The note-worthy effects of the disorder and the anisotropy inclusion, as have been unveiled below in Fig.2, are as follow: The Fermi surface topologies are observed to be distinct from the Fermi arc picture. We find that the minimally gapped portion of Fermi surface forms closed loops (hole pockets) and not arcs. However, some **k**-points of the electron pockets are found to be associated with the negative Fermi energy density of state (DOS) values when B = 0 and $\acute{\varepsilon} = 0$; for B ≠ 0 and $\acute{\varepsilon} \rightarrow 0^+$ these pockets acquire positive DOS values. The latter thus seems to justify the inclusion of the hopping anisotropy in the Hamiltonian at the theory level. For $\acute{\varepsilon} \sim 0.1$ and B =50 T, the electron pockets are found to almost disappear around ($\pm\pi$,0) and get enhanced around (0,$\pm\pi$).The negative DOS issue for some **k**-points is not found to be a serious problem, for the sum rule f (k) $=\int_{-\infty}^{+\infty} d\omega$ f($\omega$) $\rho$(k,$\omega$), where f($\omega$) is the Fermi function, does not get violated. It may be mentioned that the concept of negative DOS is not a novel one; a lucid discussion regarding the possibility of negative DOS could be found in ref. [28]. Since our observation of negative DOS is over a small region in the BZ, it has been found (and to be communicated separately) not to alter the sign of local density of electronic states (LDOS) of the sample and therefore no negative conductance is expected to be detected through the application of STS technique. As regards the values of $\acute{\varepsilon} > 0.1$, we find that the evidence of the presence of the hole pockets and the lack of that of the electron pockets around ($\pm\pi$,0) is the dominant feature in the Fermi energy DOS for this range. It appears that for meaningful calculation of all physical properties in PG state in the presence of magnetic field which requires the inclusion of scattering by imperfections, such as the present specific heat anomaly investigation, the linear thermoelectric response [29] captured by the components of conductivity tensors, etc. the range $0 \leq \acute{\varepsilon} < 0.1$ is better suited for the investigation of the under-doped cuprates.

The paper is organized as follows: In section 2 we present the mean field Hamiltonian $H_{d+id}$ in momentum space involving in-plane hopping anisotropy for PG state of under-doped YBCO with non-zero magnetic field. In section 3 we discuss the elastic scattering by impurities and relate it to the issue of the Fermi pockets as this occupies the centre-stage [30] in the magneto-quantum oscillation context. In section 4 we outline the derivation of the the chirality driven coefficient of the transverse heat current. In section 5, we present the theoretical investigation of the quantum oscillation in specific heat in the CDDW state with the inclusion of the scattering by imperfections and the hopping anisotropy in the presence of a changing magnetic field. The paper ends in section 6 with the concluding remarks.

**2 Model Hamiltonian** For a magnetic field applied in z-direction, i.e. the vector potential **A**= (0, −Bx, 0) in Landau gauge, we consider a dispersion which corresponds to a six-parameter tight-binding model:

$$\varepsilon_k (B) = - 2(t_x \cos (k_x a)+ t_y \cos (k_y a + \varphi))$$
$$+ 4t' \cos (k_x a) \cos (k_y a + \varphi/2) + \varepsilon_{LL} + \varepsilon_k^{(ADD)}, \quad (1)$$

where

$$\varepsilon_{LL} \equiv \hbar \sum_{n=0}^{\infty} (2n +1) ( \omega_c /2) + (-1)^{\sigma} (g \mu_B B /2 ) ,$$

$$n = 0, 1..., \quad (2)$$

the quantity $\varphi = (2\pi eBa^2/h)$ is the Peierls phase factor, $\varepsilon_k^{(ADD)}$ ($= -2(t_x^{(3)}\cos 2k_xa + t_y^{(3)}\cos 2k_ya) - 4t^{(4)}(\cos(k_xa)\cos(2k_ya) + \cos(k_ya)\cos(2k_xa)) + 4t^{(5)}\cos(2k_xa)\cos(2k_ya))$ is the sum of the third, fourth and fifth neighbor hopping terms and 'a' is the lattice constant (of YBCO). In Eq.(2), the first term corresponds to the Landau levels, and the second to the Zeeman splitting. The quantity $\omega_c = eB/m^*$ is the cyclotron frequency where $m^*$ is the effective mass of the electrons. In the second-quantized notation, the Hamiltonian (with index j = (1,2) below corresponding to two layers of YBCO) for the Chiral d+id density-wave state, together with the anisotropy in the hopping parameters, in the presence of magnetic field (B) can be expressed as

$$H_{d+id}(B) = \sum_{k\sigma, j=1,2} \Phi^{(j)\dagger}_{k,\sigma} E(k,B) \Phi^{(j)}_{k,\sigma} \quad (3)$$

where $\Phi^{(j)\dagger}_{k,\sigma} = (d^{\dagger(1)}_{k,\sigma}\ d^{\dagger(1)}_{k+Q,\sigma}\ d^{(2)\dagger}_{k,\sigma}\ d^{\dagger(2)}_{k+Q,\sigma})$ and $E(k,B) = [\varepsilon_k^U(B) I_{4\times 4} + \zeta_k(B)\cdot\boldsymbol{\alpha}]$. Here $\boldsymbol{\alpha} = (\alpha_1\ \alpha_2\ \alpha_3\ \alpha_4)$ with

$$\alpha_i = \begin{pmatrix} \sigma_i & 0 \\ 0 & \sigma_i \end{pmatrix} (i = 1,2,3), \alpha_4 = \begin{pmatrix} 0 & I_{2\times 2} \\ I_{2\times 2} & 0 \end{pmatrix}. \quad (4)$$

$I_{4\times 4}$ and $I_{2\times 2}$, respectively, are the 4×4 and 2×2 unit matrices; $\sigma_i$ are the Pauli matrices and $\zeta_k(B) = (-\chi_k\ -\Delta_k\ \varepsilon_k^L(B)\ t_k) -$ a four-component vector. Here the chiral order parameter[1], $D_k \exp(i\theta_k)$, is given by $D_k = (\chi_k^2 + \Delta_k^2)^{1/2}$ and $\cot\theta_k = (-\chi_k/\Delta_k)$ with

$$\chi_k = -(\chi_0/2)\sin(k_xa)\sin(k_ya), \quad (5)$$

and

$$\Delta_k = (\Delta_0(T)/2)(\cos k_xa - \cos k_ya). \quad (6)$$

Following Hackl and Vojta [29], we have introduced an anisotropy parameter έ, such that the hopping matrix elements obey $t_{x,y} = (1 \pm έ/2)t_1$ and $t'_{x,y} = (1 \pm έ/2)t'$. For έ ≠ 0, the lattice rotation symmetry is spontaneously broken. In the numerical calculations we take $t_1$ as an energy unit. We consider the simplest form of the modulation vector, $\mathbf{Q} = (\pm\pi, \pm\pi)$, which results in the opening of the gap almost in the middle of the band (see Fig.6) irrespective of the position of the chemical potential. The quantity $t_k$ is momentum conserving tunneling matrix element which for the tetragonal structure is given by $t_k = (t_0/4)(\cos k_xa - \cos k_ya)^2$. The energy eigenvalues of E(k) are $E^{(j,\nu)}(k,B) = [\varepsilon_k^U(B) + jw_k(B) + \nu t_k]$ where $\varepsilon_k^U(B) = (\varepsilon_k(B) + \varepsilon_{k+Q}(B))/2$, $\varepsilon_k^L(B) = (\varepsilon_k(B) - \varepsilon_{k+Q}(B))/2$, and $w_k(B) = [(\varepsilon_k^L(B))^2 + D_k^2]^{1/2}$. Here j is equal to ($\pm 1$) with j = +1 corresponding to the upper branch (U) and j = -1 to the lower branch(L); for a given j, $\nu = \pm 1$. With these eigenvalues, upon ignoring the Zeeman term, we find that the non-interacting Matsubara propagator is

$$G_0(k,\omega_n) = \sum_{\nu=\pm 1}\{V_k^{(U,\nu)2}(i\omega_n - E^{(U,\nu)}(k,B))^{-1}$$
$$+ V_k^{(L,\nu)2}(i\omega_n - E^{(L,\nu)}(k,B))^{-1}\}. \quad (7)$$

The quasi-particle coherence factors $(V_k^{(U,\nu)2}, V_k^{(L,\nu)2})$ are given by the expressions $V_k^{(U,\nu)2} = (1/4)[1 + (\varepsilon_k^L/w_k)]$ and $V_k^{(L,\nu)2} = (1/4)[1 - (\varepsilon_k^L/w_k)]$. The magnetic field dependence of these factors arise through $\varepsilon_k^L(B)$. According to the Luttinger rule, the chemical potential μ of the fermion number is given by the equation $(1+p) = N_s^{-1}\sum_k f(k)$ where p is the doping level, $N_s$ is the number of unit cells, and

$$f(k) = \sum_{\nu,\sigma}[V_k^{(U,\nu)2}n^{(U,\nu)}(T,k,\mu,B) + V_k^{(L,\nu)2}n^{(L,\nu)}(T,k,\mu,B)] \quad (8)$$

where $n^{(j,\nu)}(T,k,\mu,B) = (\exp\beta E_k^{(j,\nu)}(B) + 1)^{-1}$, $E_k^{(j,\nu)}(B) \equiv (E^{(j,\nu)}(k,B) - \mu)$, and $\beta = (k_BT)^{-1}$. We shall consider the value μ = −0.2130 eV and will not calculate μ by the Luttinger rule. The values of the other parameters to be used to obtain the graphical representations in this paper are $t_1$ = 0.1944 eV, $t'$ = 0.0338 eV = 0.1739 $t_1$, $t^{(3)}$ = 0.0305 eV = 0.1569 $t_1$, $t^{(4)}$ = 0.0028 eV = 0.0144 $t_1$, $t^{(5)}$ = 0.0060 eV = 0.0309 $t_1$ and $t_0$ = 0.002 $t_1$. At hole doping level ~ 10%, the pseudogap(PG) temperature $T^* \sim 155$ K. We have assumed, the experimental value $\Delta_0(T < T^*) = 0.0825$ eV $= 0.3300\ t_1$ in the vicinity of $T^*$, and $(\chi_0/\Delta_0(T<T^*))^2 = 0.0025$. We shall now consider the effect of the elastic scattering by impurities on the Fermi surface topology and search for the evidence of the existence of Fermi pockets. This is an important issue as without these pockets the Onsager relation [30] does not allow one to investigate magneto-quantum oscillations.

**3. Elastic scattering by impurities and Fermi pockets**
The impurity potential/disorder with finite range not only has drastic effects on the Fermi surface (FS) topology, but will be seen to affect the density of states at Fermi energy relevant for transport properties as well. The effect of elastic scattering by impurities involves the calculation of self-energy $\sum_{el}(k,\omega_n)$ in terms of the momentum and the Matsubara frequencies $\omega_n$. A few diagrams contributing to the self-energy are shown in Fig.1. The wiggly lines carry momentum but no energy as the scattering is assumed to be elastic. The total momentum entering each impurity vertex, depicted by a slim ellipse, is zero. We assume that impurities are alike, distributed randomly, and contribute a potential term $U(r) = \sum_j V(r - R_j)$ where $V(r - R_j)$ is the potential due to a single impurity at $R_j$. The potential term $U(r)$ is expanded in a Fourier series $U(r) = \sum_{q,j} v(q)\exp[i(r - R_j)]$. We first consider only the contribution of the Fig.1(a). We find

$$\sum_{el}(k,\omega_n) \approx N_j\sum_{k'}|v(k-k')|^2 G_0(k',\omega_n) + \sum_e$$
$$= -i\omega_n/(2|\omega_n|\tau_k) + \sum_e \approx -i/(2\tau_k) + \sum_e \quad (10)$$

where $(1/\tau_k) = 2\pi N_j \rho_0 \sum_{k'} |v(k-k')|^2$, $N_j$ is the impurity concentration, $\rho_0$ is the reciprocal band-width ($\rho_0 \sim (5t_1)^{-1}$) and $v(k-k')$ characterizes the momentum dependent impurity potential; $\sum_e$ is the part of the first order contribution to the self-energy which can be shown to be independent of k and $\omega_n$ for k near Fermi surface. One can model $v(k-k')$ by a screened exponential falloff of the form $|v(k-k')|^2 = [|v_0|^2 \kappa^2 / \{|k-k'|^2 + \kappa^2\}]$, where $\kappa^{-1}$ characterizes the range of the impurity potential, to consider the effect of the in-plane impurities as well as the out-of plane impurities. The limit $\kappa \gg |k-k'|$, which corresponds to a point-like isotropic scattering potential characterizing the in-plane impurities, will only be considered here for simplicity. Furthermore, we assume the scattering potential to be weak and choose a value $(|v_0|/t_1) \sim 0.2$ for the numerical calculation and graphical representations. We find $\sum_e / t_1 \approx 0.026$.

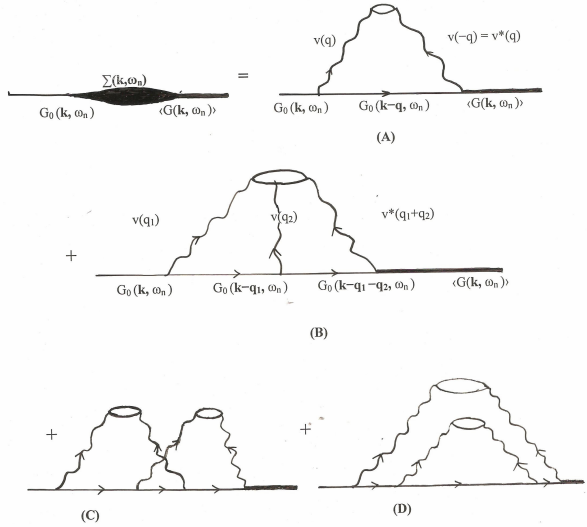

**Figure 1** A few diagrams contributing to the self-energy. The wiggly lines carry momentum but no energy. The total momentum entering each impurity vertex, depicted by a slim ellipse, is zero. We have assumed that impurities are alike, distributed randomly, and contribute a potential term $U(r) = \sum_j V(r - R_j)$ where $V(r - R_j)$ is the potential due to a single impurity at $R_j$. The potential term $U(r)$ is expanded in a Fourier series $U(r) = \sum_{q,j} v(q) \exp[i(r - R_j)]$.

Since the measurement of properties, such as conductivity, are made on a macroscopic scale in which solid appears homogeneous in terms of the density of impurities the measured quantities are usually spatial averages over regions containing large number of impurities ($N_j \gg 1$). One can model this scenario taking an ensemble average of the full Matsubara propagator $G(k,\omega_n)$ over many systems similar to one in hand containing the same average density of impurities. The ensemble average $\langle G(k,\omega_n) \rangle = G_0(k,\omega_n) / [1 - G_0(k,\omega_n) \sum_{el}(k,\omega_n)]$. In view of (10), with a magnetic field (B) as the background, after considerable algebra we obtain

$$\langle G(k,\omega_n) \rangle \approx \sum_{j=(U,L),v} V'^{(j,v)2}_k (i\omega_n - E'^{(j,v)}(k,B) + i(1/4\tau_k^{(j)}))^{-1}. \quad (11)$$

where, obviously enough, $\tau_k^{(j)}$ are the quasi-particle (QP) lifetimes. We find that

$$E'^{(j,v)}(k,B) = \varepsilon_k^U + \sum_e/2 + j R_k^{1/2} \cos(\theta_k/2) + v\, t(k) + \varepsilon_{LL},$$

$$\tau_k^{(j)-1} \approx \tau_k^{-1} \pm 4 R_k^{1/2} \sin(\theta_k/2),$$

$$R_k = [(D_k^2 + (\varepsilon_k^L + \sum_e/2)^2 - 1/16 \tau_k^2)^2 + (\varepsilon_k^L/2 \tau_k)^2]^{1/2},$$

$$\sin \theta_k = (\varepsilon_k^L / 2 \tau_k R_k),$$

$$V'^{(U,v)2}_k = (1/4)[1 + R_k^{-1/2} \{(\varepsilon_k^L + \sum_e/2 - i/4\tau_k) \exp(i\theta_k/2)\}]$$

$$V'^{(L,v)2}_k = (1/4)[1 - R_k^{-1/2} \{(\varepsilon_k^L + \sum_e/2 - i/4\tau_k) \exp(i\theta_k/2)\}]. \quad (12)$$

In the expression of $1/\tau_k^{(j)}$ above (obtained in the Born approximation for scattering), the positive sign corresponds to the upper band and the negative sign to the lower band. The expression shows that the impurity scattering leads to finite lifetime for the fermion states of definite momentum. The retarded Green's function $G_R(k,t)$ can be expressed as $G_R(k,t) = \int_{-\infty}^{+\infty} (d\omega/2\pi) \exp(-i\omega t) \langle G(k,\omega) \rangle$ where in the upper half-plane, $\langle G(k,\omega) \rangle = \sum_{j=(U,L),v} V'^{(j,v)2}_k (\omega - E'^{(j,v)}(k,B) + i(1/4\tau_k^{(j)}))^{-1}$. This leads to the result

$$G_R(k,t) \approx \sum_{j=(U,L),v} V'^{(j,v)2}_k \exp[-(t/4\tau_k^{(j)})]$$

$$\times \exp[-i(E'^{(j,v)}(k,B) t/\hbar - \pi/2)] \theta(t), \quad (13)$$

The function $G_R(k,\omega') = \int_{-\infty}^{+\infty} dt\, e^{i\omega' t} G_R(k,t)$, in turn, leads to the DOS $\rho(k,B,\omega) = -(1/2\pi^2) \text{Im}\, G_R(k,B,\omega)$. We find that $\rho(k,B,\omega)$ comprises of two parts: $\rho(k,B,\omega) = \rho_1(k,B,\omega) + \rho_2(k,B,\omega)$, where

$$\rho_1(k,B,\omega) = (t_1^{-1}/2\pi^2) \sum_{j=(U,L),v} \text{Re}\, V'^{(j,v)2}_k$$

$$\times \gamma_k^{(j)} / [(\omega/t_1 - E'^{(j,v)}(k,B)/t_1)^2 + \gamma_k^{(j)2}], \quad (14)$$

$$\rho_2(k,B,\omega) = (-t_1^{-1}/2\pi^2) \sum_{j=(U,L),v} \text{Im}\, V'^{(j,v)2}_k$$

$$\times (\omega/t_1 - E'^{(j,v)}(k,B)/t_1) / [(\omega/t_1 - E'^{(j,v)}(k,B)/t_1)^2 + \gamma_k^{(j)2}], \quad (15)$$

and $\gamma_k^{(j)} \sim \tau_k^{(j)-1}/4t_1$ (the level-broadening factors). It may be noted that $\rho_1(k, B=0, \omega=\mu)$ roughly corresponds to the so-called "maximal intensity surface" [22] of the ARPES studies provided the momentum dependence of the level broadening factors are ignored.

We now consider the contributions of all the diagrams of the type 1(a) and 1(b) involving one impurity vertex only.

The total self-energy contributions from these diagrams can be written as

$$\Sigma(k,\omega_n) = N_j \sum_q v(q) G_0(k-q,\omega_n) v(-q)$$

$$+ N_j \sum_{q,q',q''} v(q) G_0(k-q,\omega_n) v(q') G_0(k-q-q',\omega_n) v(q'')$$

$$\times \delta(q+q'+q'')$$

$$+ N_j \sum_{q,q',q'',q_1} v(q) G_0(k-q,\omega_n) v(q') G_0(k-q-q',\omega_n)$$

$$\times v(q'') G_0(k-q-q'-q'',\omega_n) v(q_1) \delta(q+q'+q''+q_1)$$

$$+ \ldots \ldots \ldots \ldots \ldots \ldots \quad (16)$$

Equation (16), in a compact form, can be written as

$$\Sigma(k,\omega_n) = N_j \sum_q v(q) G_0(k-q,\omega_n) \Gamma(k,q,\omega_n) \quad (17)$$

where the integral equation to determine $\Gamma(k,q,\omega_n)$ is given by

$$\Gamma(k,q,\omega_n) = v(-q) + \sum_{q'} v(q'-q) G_0(k-q',\omega_n) \Gamma(k,q',\omega_n). \quad (18)$$

This corresponds to the $\Gamma$-martix approximation. Upon using the optical theorem for the $\Gamma$-matrix one may write

$$\Sigma_{el}(k,\omega_n) = i \, \mathrm{Im}\, \Gamma(k,k,\omega_n) = -i\omega_n/(2|\omega_n|\acute{\Gamma}_k) \quad (19)$$

where $\acute{\Gamma}_k^{-1} = 2\pi N_j \rho_0 \sum_{k'} |\Gamma(k,k')|^2$. Thus the effect of the inclusion of all the diagrams of the type Fig.1(a), 1(b) and so on is to replace the Born approximation for scattering by the exact scattering cross-section for a single impurity. Since $G_0(k,\omega_n)$ and $v(q)$ are specified above, using Eqs. (18) and (19) one can determine $\acute{\Gamma}_k^{-1}$ in terms of $v(k)$. Thereafter, $(1/\tau_k)$ in Eq.(12) will have to be replaced by $\acute{\Gamma}_k^{-1}$. In the limit $\kappa \gg |k-k'|$, which ensures the momentum independence of $\acute{\Gamma}_k$, it is easy to see that the quantity $\gamma \equiv (\acute{\Gamma}_k^{-1}/t_1) \approx 1.6\zeta^2/(1+\zeta^2)$ where $\zeta = (2\pi\rho_0 |v_0|)$. The choice of parameters, viz. $\rho_0 \sim (5t_1)^{-1}$ and $(|v_0|/t_1) \sim 0.2$, leads to $\gamma \sim 0.1$. With the replacement $\tau_k^{-1} \rightarrow \acute{\Gamma}_k^{-1}$ in Eq.(12) one obtains the expressions for the QP lifetimes $\tau_k^{(j)}$ and the FS branches $(E^{(j,\nu)}(k,B) = \mu)$ in the $\Gamma$-martix approximation for the impurity scattering. We note that, even though $\acute{\Gamma}_k$ is found to be k-independent in the first approximation, the term $\pm 4 R_k^{1/2} \sin(\theta_k/2)$ will ensure that $\tau_k^{(j)}$ are momentum dependent and different for the upper and lower branches. The replacement also leads to $G_R(k,t)$ (see Eq.(13)) in the same approximation. The quantity $G_R(k,\omega') = \int_{-\infty}^{+\infty} dt\, e^{i\omega' t} G_R(k,t)$, in turn, leads to the Fermi energy DOS $\rho_{Fermi}(k,B) = -(1/2\pi^2)\, \mathrm{Im}\, G_R(k,\omega = \mu)$. We once again find that, in the $\Gamma$-martix approximation, $\rho(k,B,\omega=\mu)$ comprises of two parts: $\rho(k,B,\omega=\mu) = \rho_1(k,B,\omega=\mu) + \rho_2(k,B,\omega=\mu)$ where $\rho_1$ and $\rho_2$, respectively, are obtained from Eqs. (14) and (15) making the replacement $\tau_k^{-1} \rightarrow \acute{\Gamma}_k^{-1}$.

Since we do not posses a priori indication regarding the admissible numerical values of $\acute{\varepsilon}$, we shall assume $\acute{\varepsilon} \leq 0.1$. With this input we have plotted the graphs shown in Fig.2. In Fig.2(a), for example, $\rho_1(k, B=0, \omega=\mu)$ on the Brillouin zone at 10 % hole doping is depicted for the anisotropy parameter $\acute{\varepsilon} = 0$. The plot shows the usual Fermi arc feature and the outline of the 'electron pockets' centered around $[(\pm\pi, 0), (0, \pm\pi)]$. The outcome is in general

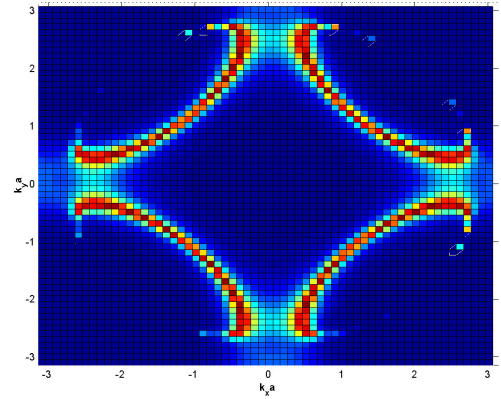

(a) B = 0 and $\acute{\varepsilon} = 0$.

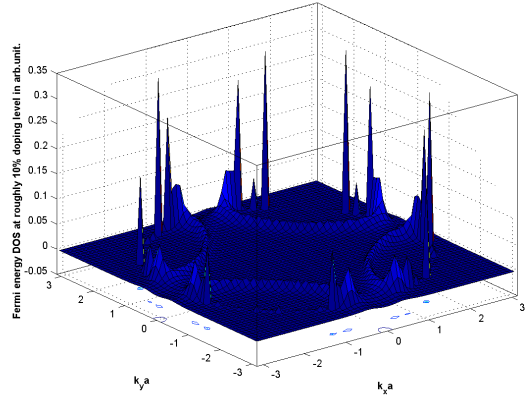

(b) B = 0 and $\acute{\varepsilon} = 0$.

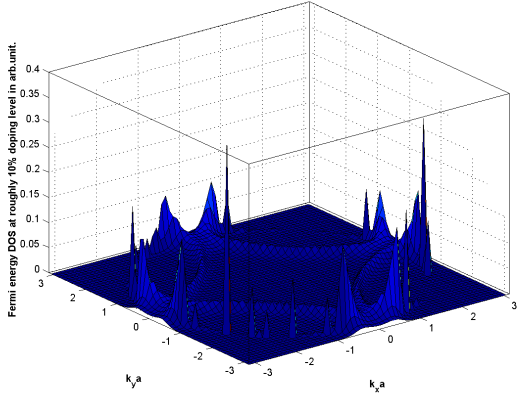

(c) B = 50 T and $\acute{\varepsilon} = 0$.

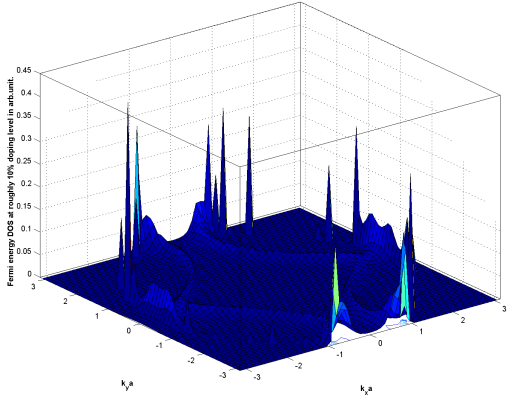

(d) B = 50 T and $\acute{\varepsilon} \sim 0.1$.

**Figure 2** (a) The contour plot of the quantity, $\rho_1(k,B=0,\omega=\mu)$, on the Brillouin zone(BZ) at 10% hole doping. The anisotropy parameter $\acute{\varepsilon}$ is assumed to be zero. The scale of the plots is from 0 to 0.14. The plots show the usual Fermi arc feature (without the hole pockets) and the outline of the electron pockets. (b) The 3-D plot of the Fermi energy DOS $\rho(k,B=0,\omega=\mu)$ on the Brillouin zone at 10 % hole doping for $\acute{\varepsilon} = 0$. The outline of the hole pockets are clearly visible. For some **k**-points on the electron pockets centered around $[(0,\pm\pi),(\pm\pi,0)]$ correspond to the negative density of states. As long as the sum rule $f(k) = \int_{-\infty}^{+\infty} d\omega\, f(\omega)\, \rho(k,\omega)$ is not violated one has no reason to worry about the negative DOS. (c) The 3-D plot of the Fermi energy DOS $\rho(k,B=50\,T,\omega=\mu)$ for $\acute{\varepsilon} = 0$. The Fermi pockets correspond to the positive density of states. (d) The 3-D plot of the Fermi energy DOS $\rho(k,B=50\,T,\omega=\mu)$ for $\acute{\varepsilon} \sim 0.1$. The electron pockets almost disappear around $(\pm\pi,0)$ and get enhanced around $(0,\pm\pi)$. In fact, those around $(0,\pm\pi)$ correspond to patches slightly bulging upward at the boundary. The outline of the hole pockets are clearly visible. The numerical values of the other parameters used to obtain these graphical representations are given in the text above.

agreement with the experimental data [20, 21, 22, 23, 24, 25, 26] on $YBa_2Cu_3O_{6+y}$ and other hole-doped high temperature superconductors. The ARPES data, as already mentioned, have not exhibited the signature of the Fermi pockets so far. However, as we see in Fig.2(b), a plot of $\rho(k,B=0,\omega=\mu)$ yields the missing hole pockets for $\acute{\varepsilon} = 0$. The reason, needless to say, is the inclusion of the crucial term $\rho_2(k,B=0,\omega=\mu)$ [31] discussed in section 1. It must be noted that some **k**-points in the electron pockets in Fig.2(b) centered around $[(0,\pm\pi), (\pm\pi,0)]$ correspond to the negative density of states(DOS). As long as the sum rule $f(k) = \int_{-\infty}^{+\infty} d\omega\, f(\omega)\, \rho(k,\omega)$ is not violated one has no reason to worry about the negative DOS. In Fig.2(c) we find that reasonably well-formed electron pockets around $[(\pm\pi,0), (0,\pm\pi)]$ do appear for magnetic field B ~ 50 Tesla and the anisotropy parameter $\acute{\varepsilon} = 0$ in the plot of the $\rho_{Fermi}(k,B)$ with positive DOS. Evidently, while in the absence of magnetic field some **k**-points on the electron pockets correspond to negative DOS, once the field is turned on and it attains a value to be able to trigger Landau level quantization and the greater coherence(see the discussion below and ref. [32]) of the quasi-particles, the pockets around $[(\pm\pi,0), (0,\pm\pi)]$ acquire positive DOS values. However, as shown in Fig.2(d), for $\acute{\varepsilon} \sim 0.1$ the electron pockets around $(\pm\pi,0)$ practically disappear and those around $(0,\pm\pi)$ become quite visible.

Since the exercise above has not been able to indicate precisely as to what should be the upper limit of $\acute{\varepsilon}$ for the electron pockets around $(\pm\pi,0)$ to remain intact in the presence of magnetic field, we now attempt to obtain this value through the investigation of an altogether different quantity, viz. the quasi-particle lifetime (QPLT). The significantly higher lifetime of the upper band (electron-like) quasi-particles compared to that of the lower band (hole-like) quasi-particle in the presence of a magnetic field (~ 50 T) in the anti-nodal and the nodal regions on the Brillouin zone follows in our investigation for $\acute{\varepsilon} = 0$ (see Figs.3(a) and 3(b)). The reason is the difference of the effective masses $(1/m_{x,y}^{(j=U,L)} = \hbar^{-2}(d^2\, E^{(j=U,L)}(k,B)/dk_{x,y}^2)$, between the CDDW quasi-particles of the upper and lower bands (see Fig.3(c)). Furthermore, the upper band QPLT in the nodal region is also found to be significantly higher than that in the anti-nodal region. This explains the so-called 'nodal-antinodal dichotomy'. However, for $\acute{\varepsilon}$ higher than 0.03, the opposite scenario presents itself. Thus we shall settle for the admissible upper limit of $\acute{\varepsilon}$ as 0.03. The missing hole pockets and the electron pockets centered around $(\pm\pi,0)$ (with positive Fermi energy DOS) remain intact for $\acute{\varepsilon} < 0.03$. It must emphasized that the DDW/CDDW order is assumed to co-exist/co-operate with the spontaneously broken lattice rotation symmetry for $\acute{\varepsilon} \neq 0$ here. In section 5 we have made an effort to relate these findings with the magneto-quantum oscillations in the specific heat.

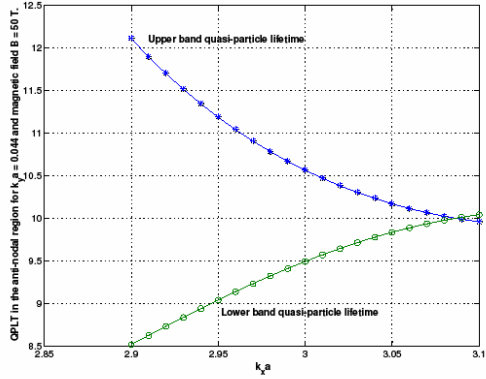

(a)

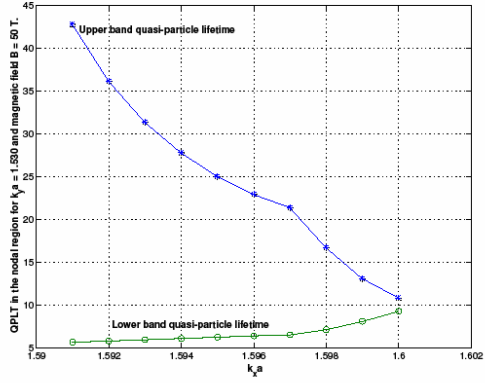

(b)

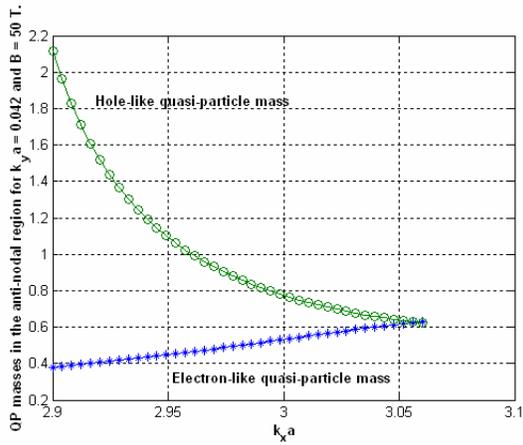

(c)

**Figure 3** The plot of the upper and the lower band quasi-particle lifetime (QPLT) in the anti-nodal and the nodal regions at 10% hole doping in the presence of B = 50 T. The anisotropy parameter $\acute{\varepsilon}$ is assumed to be zero. (a) Here $k_y a = 0.044$ and $2.900 \leq k_x a \leq 3.100$. (b) Here $k_y a = 1.530$ and $1.591 \leq k_x a \leq 1.600$. (c) The plot of the upper and the lower band quasi-particle masses in the anti-nodal region at 10% hole doping for $\acute{\varepsilon} = 0$. Here $k_y a = 0.042$ and $2.900 \leq k_x a \leq 3.060$.

## 4. Entropy density and chirality induced Nernst signal

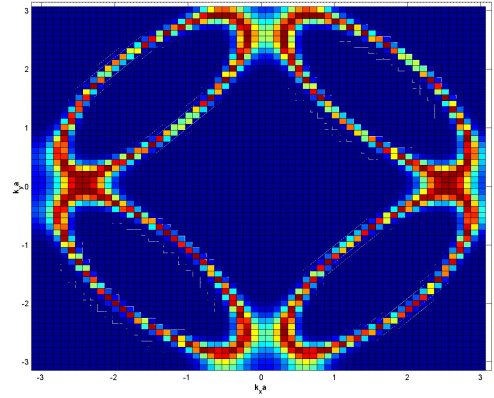

(a)

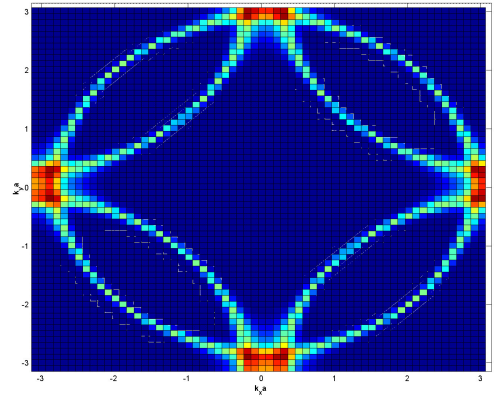

(b)

**Figure 4.** The contour plots of the entropy densities $s_{PG}(\mathbf{k},B)$ and $s_N(\mathbf{k})$ on the Brillouin zone. (a) This plot corresponds to $s_{PG}(\mathbf{k},B)$ for B = 50 T and $\acute{\varepsilon} = 0.03$. The scale of the plot is from 0 to 0.7. The clear evidence of the existence of the hole pockets and emerging electron pockets around $(0,\pm\pi)$ are present in the plot. (b) This plot corresponds to $s_N(\mathbf{k})$ for B = 0 and $\acute{\varepsilon} = 0$. The scale of the plots is from 0 to 1.4. The scale of the plots are indicative of $\sum \Delta s(\mathbf{k})$ being negative. The clear evidence of the existence of the electron pockets are present in $s_N(\mathbf{k})$.

The thermodynamic potential is given by the expression $\Omega(B) = \Omega_0(B) - 2(\beta N_s)^{-1} \sum_{k,j(=U,L),\nu} \{\ln\cosh(\beta E_k^{\prime(j,\nu)}(B)/2)\}$ where $E_k^{\prime(j,\nu)}(B) \equiv (E^{\prime(j,\nu)}(k,B) - \mu)$, and $\Omega_0(B) = N_s^{-1} \sum_{k,j(=U,L),\nu} E_k^{(j,\nu)}$. The dimensionless entropy per unit cell is given by $S = \beta^2 (\partial \Omega/\partial \beta)$. For the pseudo-gapped (PG) phase ($T < T^*$) and the normal phase ($T > T^*$) the entropy expressions are $S_{PG}(B) = 2N_s^{-1} \sum_k s_{PG}(k,B)$ and $S_N = 2N_s^{-1}\sum_k s_N(k)$, respectively, where

$$s_{PG}(k,B) = \sum_{j(=U,L),\nu} [\ln(\exp(-\beta E_k^{\prime(j,\nu)}(B)) + 1)$$
$$+ (\beta E_k^{\prime(j,\nu)}(B) + \beta^2 (\partial E_k^{\prime(j,\nu)}(B)/\partial \beta))$$
$$\times (\exp(\beta E_k^{\prime(j,\nu)}(B)) + 1)^{-1}], \quad (20)$$

$$s_N(k) = \sum_{j=1,2} [\ln(\exp(-\beta \varepsilon_j(k)) + 1)$$
$$+ (\beta \varepsilon_j(k) + \beta^2 (\partial \varepsilon_j(k)/\partial \beta)) (\exp(\beta \varepsilon_j(k) + 1)^{-1}], \quad (21)$$

$\varepsilon_1(k) = \varepsilon_k$, and $\varepsilon_2(k) = \varepsilon_{k+Q}$. To calculate $s_{PG}(k, B)$ and $s_N(k)$, respectively, we take the temperature equal to 150 K and 160 K. Furthermore, we assume that at $T = T^*$ the onset of the CDDW ordering and the hopping anisotropy take place simultaneously. The contour plots of $s_{PG}(k,B)$ and $s_N(k)$ on the Brillouin zone at 10% hole doping is shown in Fig.4. We obtain the corroboration of the facts revealed in Fig.3: The signature of the emerging electron pockets around $(0,\pm\pi)$ are present in $s_{PG}(k, B = 50$ T$)$ for, say, $\acute{\varepsilon} = 0.03$. Now as mentioned in section 1, the identification of the pseudo-gapped phase with chiral DDW ordered phase was necessary to explain [5] the findings of Kapitulnik et al.[11]. It is imperative that we discuss the effects of the chirality[1,5,33] in this phase. Since the anomalous Nernst signal(ANS) is linked to chirality as well as entropy, in what follows we wish to discuss ANS briefly.

In the presence of an external electric field **E** along y-direction (and magnetic field **B** =0), the transverse heat current $J_x$ in the x-direction is given by the relation $J_x = (T S_{xy} E_y)$ where the coefficient $S_{xy}$ is defined below. The reason for nonzero $J_x$ is that, in the presence of chirality, the carriers acquire an anomalous velocity $v_a$ given by $\hbar v_a = e\, E \times \Omega^{(a)}(k)$ where for $\acute{\varepsilon} = 0$

$$\Omega^{(a=1,2)}(k_x, k_y) = \pm\, ta^2 w_k^{-3} \chi_0 \Delta_0$$
$$\times (\sin^2 k_y a + \sin^2 k_x a \cos^2 k_y a) \quad (22)$$

are the Berry curvatures(BCs) having opposite signs with nonzero component only in the z-direction (see Refs.[1,2,5]). However, for $\acute{\varepsilon} \neq 0$, we have

$$\Omega^{(a=1,2)}(k_x, k_y) = \pm\, a^2 (\chi_0 \Delta_0/t_1^2)(g_1+g_2)/(g_3 + g_4)^{3/2} \quad (23)$$

where

$$g_1 = 0.5(\cos k_x a - \cos k_y a) \times \{(1 - \acute{\varepsilon}/2) \cos k_x a \sin^2 k_y a$$
$$- (1 + \acute{\varepsilon}/2) \sin^2 k_x a \cos k_y a\} + \sin^2 k_x a \sin^2 k_y a, \quad (24)$$

$$g_2 = 0.5(\cos k_x a \sin^2 k_y a + \sin^2 k_x a \cos k_y a)$$
$$\times \{(1 + \acute{\varepsilon}/2) \cos k_x a + (1 - \acute{\varepsilon}/2) \cos k_y a\}, \quad (25)$$

$$g_3 = 4\{(1 + \acute{\varepsilon}/2) \cos k_x a + (1 - \acute{\varepsilon}/2) \cos k_y a\}^2, \quad (26)$$

$$g_4 = \{(\chi_0/2t_1)^2 \sin^2 k_x a \sin^2 k_y a$$
$$+ (\Delta_0/2t_1)^2 (\cos k_x a - \cos k_y a)^2\}. \quad (27)$$

Upon multiplying $v_a$ by the entropy density for B =0 we obtain the coefficient $S_{xy}$ for the transverse heat current:

$$S_{xy}(T<T^*) = (e/\hbar N_s a^2) \sum_{k,a=1,2} \Omega^{(a)}(k)\, k_B\, s_{PG}^{(a)}(k). \quad (28)$$

Here $\Omega^{(1)}(k)(> 0)$ is multiplied with $s_{PG}(k)$ corresponding to the upper (U) branch and $\Omega^{(2)}(k)(< 0)$ to that for the lower (L) branch. The BCs peak on the hole pockets and much less prominently elsewhere along the boundary of RBZ [1,5]. We find from Fig.4 that $s_{PG}(k)$ peaks are on the boundary of the hole pockets only. It follows that the points common to the RBZ boundary and the hole pockets will be the major contributor towards the momentum integral, $\sum_{k,a=1,2} \Omega^{(a)}(k)\, s_{PG}^{(a)}(k)$, in Eq.(28); the electron pocket contributions will be far too less. The statements to this effect could be found in section 3 of ref.[1]. The graphical representation in Fig.5 here also underscores this fact. We find that the peaks in $\sum_{a=1,2} \Omega^{(a)}(k)\, s_{PG}^{(a)}(k)$ occur at $(\pm\pi(1-\delta), \pm\pi\delta), (\pm\pi\delta, \pm\pi(1-\delta))$ with $\delta \sim 0.24$ on the boundary of RBZ.

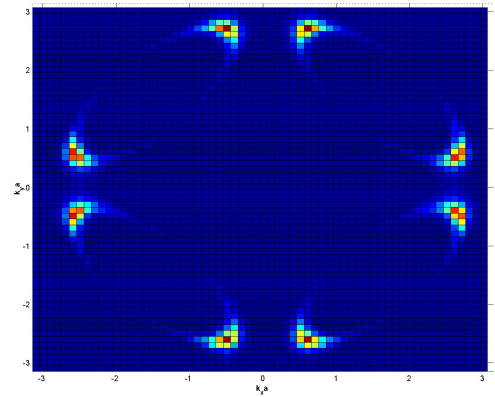

**Figure 5** A contour plot of the integrand (in Eq.(28)) $\sum_{a=1,2} \Omega^{(a)}(\mathbf{k}) \, s_{PG}^{(a)}(\mathbf{k}, B = 0)$ on the Brillouin zone(BZ) at 10% hole doping. We have taken $(\chi_0/\Delta_0 (T<T^*))^2 = 0.0025$ to obtain this plot. The scale of the plot is from $-0.01$ to $0.07$. We find that the peaks are on the boundary of the hole pockets (the peaks occur at $(\pm\pi(1-\delta), \pm\pi\delta), (\pm\pi\delta, \pm\pi(1-\delta))$ with $\delta \sim 0.24$). Elsewhere on the BZ, the integral is practically zero.

## 5. Quantum oscillations in specific heat

We shall now discuss the specific heat oscillation at low temperature in the presence of a changing magnetic field in the CDDW state. The specific heat $(C = -\beta\,(\partial S_{PG}/\partial\beta))$ for $B \ne 0$ is easily obtained from the entropy $S_{PG}(B)$. We find

$$C \approx 2k_B N_s^{-1} \sum_{k,j,\nu} (\beta E'^{(j,\nu)}_k)^2 \exp(\beta E'^{(j,\nu)}_k)(\exp(\beta E'^{(j,\nu)}_k)+1)^{-2}. \quad (29)$$

We have ignored the temperature dependence of the chemical potential above. Upon using (14) and (15) in (29) we find that the specific heat at a given doping level is given by $C \approx \gamma(B)\,T$, where the specific heat coefficient for $B \ne 0$ may be expressed as

$$\gamma(B) \approx (2k_B^2/\pi\,\hbar\,\omega_c) \sum_{j,\nu,\sigma} \int d\mathbf{k}\, Q(B,\mathbf{k}) \quad (30)$$

where

$$Q(B,\mathbf{k}) = \int_0^{+\infty} dx \int_0^{\infty} dm\, \cos(2mx/\beta\hbar\omega_c)\{x^2 e^x/(e^x+1)^2\}$$
$$\times [\, V''^{(j,\nu)2}_k \exp(-2m\gamma_k^{(j)} t_1/\hbar\omega_c)$$
$$\times \cos\{m(2E'^{(j,\nu)}_k/(\hbar\omega_c)+\theta)\}\,], \quad (31)$$

$$V''^{(j,\nu)2}_k = \mathrm{Re}V'^{(j,\nu)2}_k + \mathrm{Im}V'^{(j,\nu)2}_k$$
$$\times\{(2E'^{(j,\nu)}_k/(\hbar\omega_c)+\theta)/(2\gamma_k^{(j)} t_1/\hbar\omega_c)\}, \quad (32)$$

$\theta = \sum_{n=0}^{\infty}(2n+1)$ and $\int d\mathbf{k} \to \int_{-\pi}^{+\pi}(d(k_xa)/2\pi \int_{-\pi}^{+\pi}(d(k_ya)/2\pi$; the Zeeman term is assumed to be insignificant. To arrive at Eq.(31) we have made use of the integral $\int_0^{\infty} dm\, e^{-am} \cos(bm) = a/(|b|^2+a^2)$, $a > 0$. Equation (31) indicates that apparently the origin of the approximate $(1/B)$-oscillations are the upper and the lower branches of the excitation spectrum (or the electron and the hole pockets, respectively) albeit with different Dingle factors and frequency. The low-frequency conductivity oscillations in $YBa_2Cu_3O_{6.5}$ and $YBa_2Cu_4O_8$ have been observed on the background of a negative Hall coefficient $R_H$ at low temperature and magnetic field $\sim 50$ T (ref.[14,15]). The implication is that the electron pocket is the origin of the main frequency of oscillations at this value of the field. We show this below in the theoretical investigation of ours.

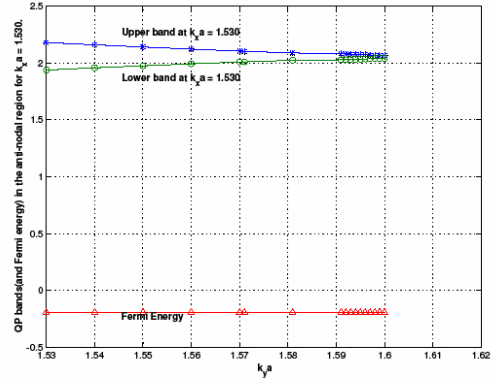

(b)

**Figure 6** The plots of the quasi-particle upper and the lower bands in the anti-nodal and the nodal regions and the chemical potential at 10% hole doping in the presence of B = 50 T. The anisotropy parameter $\acute{\epsilon}$ is assumed to be 0.03.(a) Here $k_x a = 0.044$ and $2.900 \le k_y a \le 3.100$. (b) Here $k_x a = 1.530$ and $1.530 \le k_y a \le 1.620$.

In order to estimate the oscillation frequencies($F_U$ and $F_L$), we make the following approximation for the anti-nodal region and $\acute{\epsilon} = 0.03$ ( see Figs.6(a) and (b))

$$E'^{(U,\nu)}_k \approx E'^{(U,\nu)}_k \big|_{\mathbf{k}=(0.044,\pm 2.950)} \equiv \tilde{\bar{E}}_U,$$

$$E'^{(L,\nu)}_k \approx E'^{(L,\nu)}_k \big|_{\mathbf{k}=(0.044,\pm 2.950)} \equiv \tilde{\bar{E}}_L$$

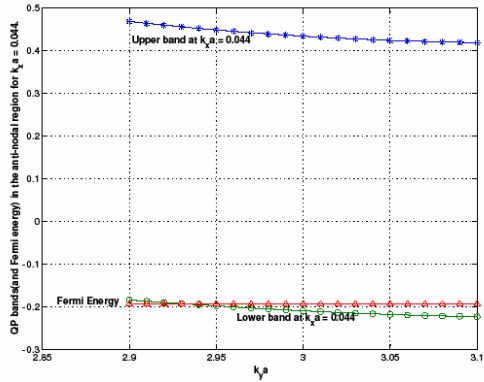

(a)

which simplifies our task greatly. We obtain the magnitudes of $\tilde{\bar{E}}_U$ and $\tilde{\bar{E}}_L$, respectively, as 0.0871eV and 0.0385 eV. Similarly, $E'^{(U,\nu)}_k \big|_{\mathbf{k}=(1.530,\pm 1.591)} \approx E'^{(L,\nu)}_k \big|_{\mathbf{k}=(1.530,\pm 1.591)} \equiv \tilde{\bar{E}}_L' = 0.3985$ eV. Upon equating $(2\tilde{\bar{E}}_U/\hbar\omega_c)$ and $(2\tilde{\bar{E}}_L'$

$/\hbar\omega_c$) with ($2\pi F_U/B$) and ($2\pi F_L/B$), respectively, we find that the frequencies are $F_U \sim 240$ T and $F_L \sim 1100$ T. Using the values of $F_U$ ($\approx 240$ T) and $F_L$ ($\sim 1100$ T) and the Onsager relation [30] we find that the hole pocket covers an area approximately 4.6 times bigger than that of the electron pocket. The key inputs in the investigation above are DOS given by Eqs. (14) and (15) and the single-particle excitation spectrum given by Eq.(12) involving the Landau level quantization; the chirality aspect plays a minor role as these oscillations are also possible in the pure d-density wave state[4]. These periodic quantum oscillations are in principle observable in all solid state properties of YBCO in the pseudo-gapped state.

**6. Concluding remarks** The well-known theoretical developments, such as the symmetry-constrained variational procedure of Wu et al. [34] yielding the generalization of BCS paradigm, the dynamical mean field theory (DMFT)[7,8],etc., may require in future a revisit of the problem of the quantum oscillations with a new perspective. Particularly, the development of DMFT and its cluster extensions provide new path to investigate strongly correlated systems; the DMFT study of superconductivity near the Mott transition establishes the remarkable coexistence of a superconducting gap, stemming from the anomalous self-energy, with a pseudo-gap stemming from the normal self-energy. This theory also leads to the generation of the Fermi arc behavior of the spectral function [7,8]. A complete structure of DMFT compatible with the findings of Doiron-Leyraud et al. [14,15], however, is yet to emerge.

One would like to say a few words regarding why the choice of six-parameter tight-binding model has been made in Eq. (1). If one proceeds with the usual four-parameter or three-parameter dispersions [1,3,4] the plot of the Fermi energy DOS for $έ = 0$ yields the outline of the electron and the hole pockets. However, the electron pockets are found to correspond to the negative density of states entirely. Though as long as the sum rule (see section 1) is not violated one has no reason to worry about the negative DOS, a very large sub-set of the set of k-points in the Brillouin zone corresponding to the negative Fermi energy DOS has been regarded as an unwelcome feature by us. The present choice, as seen above, leads to a more reasonable scenario.

In conclusion, we note that the computation of the correction to the quantum oscillations due to the Berry phase is an important future task. The Dirac cone like feature in the quasi-particle lower band is yet another issue ahead of us which needs to be examined. We note that the inclusion of the elastic scattering by impurities though has led to a clearer understanding, of the Fermi surface topology in the presence of a magnetic field at the semi-phenomenological level, the further examination of the single-particle excitation spectrum of the system in a fully self-consistent approximation framework is necessary to impart a comprehensive microscopic basis to the findings presented. Finally, we hope that our results, viz. the one relating to the reconstructed Fermi surface and the other to the electronic specific heat anomaly, will persuade researchers to look for them. The observation of the latter is quite a difficult proposition though, in the heat capacity measurements [35], as the dominant phononic contribution is expected to overshadow this anomaly.